\documentclass[showpacs,preprintnumbers,amsmath,amssymb,nofootinbib]{revtex4}

\usepackage{graphicx}
\usepackage{bm}
\usepackage{appendix}
\usepackage{epstopdf}
\usepackage{float}
\usepackage{dcolumn}
\usepackage{bm}
\usepackage{mathrsfs}
\usepackage{amsmath}

\begin{document}

\title{Newtonian Cosmology with a Quantum Bounce}
\author{P. Bargue\~no}
\email{p.bargueno@uniandes.edu.co}
\author{S. Bravo Medina}
\email{s.bravo58@uniandes.edu.co }
\author{M. Nowakowski}
\email{mnowakos@uniandes.edu.co}
\affiliation{%
Departamento de Fisica,\\ Universidad de los Andes, Cra.1E
No.18A-10, Bogota, Colombia
}
\author{D. Batic}
\email{davide.batic@uwimona.edu.jm}
\affiliation{%
Department of Mathematics, University of West Indies,
Kingston 6, Jamaica}

\date{\today}

\begin{abstract}
It has been known for some time that the cosmological Friedmann equation deduced from General
Relativity can be also obtained within the Newtonian framework under certain assumptions.
We use this result together with quantum corrections to the Newtonian potentials
to derive a set a  of quantum corrected Friedmann equations. We examine the behavior of the solutions of these modified 
cosmological equations paying special attention to the sign of the quantum corrections. We find different
quantum effects crucially depending on this sign. One such a solution displays a qualitative resemblance to
other quantum models like Loop Quantum Gravity or non-commutative geometry.    
\end{abstract}

\pacs{Valid PACS appear here}
\maketitle
\section{Introduction}
It must have come as a surprise to the physics community when McCrea and Milne \cite{McCrea}
derived the
cosmological Friedmann equations known from General Relativity from Newtonian mechanics
assuming the existence of expansion. The interest in this derivation has persisted over years \cite{Callan,Tipler,Milne,Layzer,McCrea2,McCrea3,Thatcher,Jordan,Wells} paying attention
to refine the Newtonian set up and conclusions. In this paper we go one step further and put forward the question of
what kind of modified Friedmann equations would emerge if we include quantum corrections to the Newtonian potential which, of course, 
is one of the main ingredients in the Newtonian derivation of the Friedmann equations. Such corrections have been known for some time
\cite{Donoghue1994,HamberLiu,Muzinich,Akhundov,KK,PRD2003a,PRD2003,Ross-Holstein,Kirilin,Haranas-Mioc,Donoghue2012,BDH2015,Rovelli,Woodard,Radkowski}. As we will show below, it is straightforward to repeat the McCrea-Milne derivation
including these quantum corrections and to arrive at the modified Friedmann equations with terms proportional to $\hbar$. 
Depending on the sign of the quantum corrections (and also on the equation of state) different quantum effects emerge with one of them
resembling qualitatively those of other quantum models.  This similarity consists in the behavior of the scale factor $R$ describing a universe
which has been contracting in the past, reaching a minimal value of $R$ and expanding again after the bounce.
For the other sign of the quantum effect, the behavior is qualitatively different as the universe spontaneously appears at $R_{min}$ close to the Planck length
and starts expanding from this point. The primary expansion is accelerated reminding us of inflation. We do not claim that our modified Friedmann equations
give necessarily the correct description of a quantum universe, but it is certainly worthwhile to consider them. For one they give the right Friedmann equation
when no quantum corrections are included and as such could contain the right clues and hindsights when we include the latter. Secondly, we think it is timely
to make venture one step more in the area of Newtonian cosmologies. 
     
The paper is organized as follows. In the next section we give a brief account of the derivation of the standard
Friedmann equation within the Newtonian framework.  Next we introduce the quantum corrections and
derive the modified Friedmann equations. In section IV we study the behavior of these new cosmological
equations varying the sign of the quantum corrections and choosing different equations of state.

\section{Friedmann Equations from Newtonian Dynamics}
There exist different derivations of the cosmological Friedmann equation from the Newtonian dynamics
\cite{McCrea,Callan,Tipler,Milne,Layzer,McCrea2,McCrea3,Thatcher, Jordan, Wells} and although conceptually such derivations
differ \cite{Tipler, Wells}, in the end they all arrive at the same Friedmann equations. We therefore take here the simplest
and original point of view which starts by taking into account the expansion of the universe. 
This is
done by writing
\begin{equation} \label{expansion}
\frac{dR}{dt}=HR
\end{equation}
where $H$ is the Hubble parameter and $R$ is a measure of distance.
The next step considers the total energy for an object (say, a galaxy) of mass $m$ which reads
\begin{equation} \label{energy}
E=\frac{1}{2}m \left(\frac{dR}{dt} \right)^{2}-\frac{GMm}{R}
\end{equation}
Writing the mass inside the sphere of radius $R$ as 
$M=\frac{4}{3}\pi R^{3}\rho$ where $\rho$ is the density of the universe, equation (\ref{energy}) takes the form
\begin{equation} \label{H0}
\frac{2E}{mR^{2}}=H^{2}-\frac{8}{3}\pi G \rho
\end{equation}
Since $E$ and $m$ are constants we define $k \equiv \frac{2E}{m}$ and obtain the first Friedmann equation
\begin{equation} \label{H1}
H^{2}=\frac{8\pi G}{3}\rho - \frac{k^{2}}{R^{2}}
\end{equation}

To get the second Friedmann equation the argument goes as follows: when the volume $V$ of the universe expands by  
$dV$, the pressure does work equal to $pdV$ which decreases the energy in $V$ by that amount. Using energy mass equivalence we
one obtains
\begin{equation} \label{work}
d\left( \rho \frac{4}{3}\pi R^{3}\right)=-pd\left(\frac{4}{3}\pi R^{3} \right)
\end{equation}
On the other hand this is the well known conservation law which can be cast in the convenient form
\begin{equation} \label{work2}
R\frac{d\rho}{dt}+3(\rho+p)\frac{dR}{dt}=0
\end{equation}
If we write the first Friedmann equation as
\begin{equation}
\left(\frac{dR}{dt} \right)^{2}=\frac{8\pi G}{3}\rho R^{2}-kR(t_{1})^{2}
\end{equation}
By taking a derivative of this equation and replacing $R\frac{d\rho}{dt}$ from the conservation equation one arrives at the second Firedmann equation
\begin{equation} \label{second}
\frac{d^{2}R}{dt^{2}}=-\frac{4\pi G}{3}(\rho+3p)R
\end{equation}
From our point of view the crucial ingredient is how the Newtonian potential enters the derivation. Quantum corrections to the latter
are known and it makes some sense to try to re-derive the cosmological equations by taking this correction into account. 

\section{Quantum Corrected Friedmann Equations}
Several authors have obtained $\hbar$ corrections to the Newtonian potential by taking gravity as an effective theory and
performing one-loop graviton calculations \cite{Donoghue1994,HamberLiu,Muzinich,Akhundov,KK,PRD2003a,PRD2003,Ross-Holstein,Kirilin,Haranas-Mioc,Donoghue2012,BDH2015,Rovelli, Woodard,Radkowski,CapperDuff,Burgess,Duff}. 
\begin{equation} \label{hbarN}
\phi (r) =-\frac{GM_1M_2}{r}\left[1-\gamma_{q} \frac{G\hbar}{r^{2}c^{3}} \right]
\end{equation}
Sometimes the results of the quantum corrected Newtonian potential is given in a different forms
\begin{equation} \label{N1}
\Phi(r)=-\frac{G M_{1}M_{2}}{r}\left[ 1 +\lambda \frac{G(M_{1}+M_{2})}{rc^{2}}-\tilde{\gamma} \frac{G \hbar}{r^{2}c^{3}}+\ldots \right]
\end{equation}
where the $\lambda$ and $\tilde{\gamma}$ are parameters which take different values 
depending on the author(s).
Partly, we can attribute the reason for these discrepancies to the 
precise coordinate definition used in the calculation \cite{Burgess}. 
The question about the ambiguity of this potential due to the lack of clarity on the coordinates has
also been risen in some related articles \cite{PRD2003a}, \cite{Burgess}, \cite{Gauge}.
It is argued that a redefinition $r\rightarrow r'=r(1+aGM/r)$ would change the 
parameter $\lambda$ without affecting the observables. The general consensus is
that we can write the corrected potential as given in equation (\ref{hbarN}).
The aforementioned re-parametrization freedom still cannot account for all the
discrepancies of the different $\gamma_q$'s found in the literature. A
number of errors have been identified \cite{PRD2003a}, \cite{Donoghue1994}, but
it is not clear if this accounts for all the different values available.
It is therefore fair to list some of the results (see Table \ref{table:nonlin}).
In the table we have collected the different values for $\gamma_{q}$ which also vary in sign (we will see
that the sign plays the most important role in the cosmology derived from these corrections).
\\
\\

\begin{table}
\caption{Different values of $\gamma_{q}$ found in the literature.}
\begin{center}
\begin{tabular}{ |c|c| } 
 \hline
 \textbf{(Year) Reference} & $\gamma_{q}$ 
 \\
 \hline 
 (1994) \cite{Donoghue1994} &   \parbox[c][4ex]{6ex}{\centering $\frac{127}{30\pi^{2}}$}\\ 
 \hline 
 (1995) \cite{HamberLiu} & \parbox[c][4ex]{6ex}{\centering $\frac{122}{15\pi}$}\\
 \hline
 (1995) \cite{Muzinich} & \parbox[c][4ex]{6ex}{\centering $-\frac{17}{20\pi}$}\\
 \hline
 (1998) \cite{Akhundov} & \parbox[c][4ex]{6ex}{\centering $\frac{107}{10\pi^{2}}$}\\
 \hline
 (2002) \cite{KK} & \parbox[c][4ex]{6ex}{\centering $-\frac{121}{10\pi}$}\\
 \hline
 (2003) \cite{PRD2003a} & \parbox[c][4ex]{6ex}{\centering $-\frac{41}{10\pi}$}\\
 \hline
 (2003) \cite{PRD2003} & \parbox[c][4ex]{6ex}{\centering $-\frac{167}{30\pi}$}\\
 \hline
 (2007) \cite{Ross-Holstein} & \parbox[c][4ex]{6ex}{\centering $-\frac{41}{10}$ } \\ 
 \hline
(2007) \cite{Kirilin} & \parbox[c][4ex]{6ex}{\centering $\frac{107}{30\pi}$ } \\ 
 \hline
 (2002) \cite{Haranas-Mioc} & \parbox[c][4ex]{6ex}{\centering $\frac{122}{15\pi}$}\\
 \hline
(2012) \cite{Donoghue2012} & \parbox[c][4ex]{6ex}{\centering $-\frac{41}{10\pi}$}\\
 \hline
 (2015) \cite{BDH2015} & \parbox[c][4ex]{6ex}{\centering $-\frac{41}{10}$}\\
 \hline
\end{tabular}
\end{center} 
\label{table:nonlin} 
\end{table}
Having established the quantum correction we can proceed as before.
The total energy receives a new contribution due to the quantum correction
in the Newtonian potential, i..e,
\begin{equation}\label{energy2}
E=\frac{1}{2}m\left(\frac{dR}{dt} \right)^{2}-\frac{GMm}{R}+\gamma_{q} \frac{G^{2}\hbar M m}{R^{3}c^{3}}
\end{equation}
Introducing again the density $\rho$ and the Planck length $l_{p}=\sqrt{\frac{G\hbar}{c^{3}}}$ the above
equation is equivalent to
\begin{equation} 
\frac{2E}{m R^{2}}=\frac{1}{R^{2}}\left(\frac{dR}{dt} \right)^{2}-\frac{8}{3}\pi G \rho +\frac{8}{3}\pi G \rho \frac{l_{p}^{2}\gamma_{q}}{R^{2}}
\end{equation}
With the help of (\ref{expansion}) the first Friedmann equation with an $\hbar$-correction can be given as
\begin{equation} \label{Fh1}
H^{2}=\frac{8\pi G}{3}\rho -\frac{8\pi G}{3}\rho \frac{l_{p}^{2}\gamma_{q}}{R^{2}}-\frac{k}{R^{2}}
\end{equation}
The second corrected Friedmann equation follows form the faat that the conservation law (\ref{work2}) remains unchanged. 
We can procced as before to obtain
\begin{equation} \label{Fh2}
\frac{d^{2}R}{dt^{2}}=-\frac{4\pi G}{3}(\rho+3p)R+4\pi G l_{p}^{2}\gamma_{q} \frac{(\rho+p)}{R}
\end{equation}
We consider the equations (\ref{Fh1}) and (\ref{Fh2}) as the quantum corrected Friedmann equations derived
withing the framework of Newtonian mechanics. We will show below that they imply a quantum bounce or in
other words the initial singularity at $R=0$ is avoided. 

For the sake of comparison with other models and a better understanding of similarities and differences
between the standard Friemdann equations and the equation (\ref{Fh1}) and (\ref{Fh2})
we can re-cast the latter in different forms. 
By re-introducing the cosmological constant $\Lambda$ and taking an initially flat universe with $k=0$.
We have then
\begin{equation} \label{Fh1L}
H^{2}=\frac{8\pi G}{3}\rho + \frac{\Lambda}{3}-\frac{8\pi G}{3}\rho \frac{l_{p}^{2}\gamma_{q}}{R^{2}}
\end{equation}
Making use of the standard definitions $\rho_{crit}(t)=\frac{8\pi G}{3H^{2}}$ (provided $H$ is non-zero)  and $\rho_{vac}=\frac{\Lambda}{8\pi G}$
the first Friedmann equation (with the cosmological constant and the $\hbar$ corrections) is simply
\begin{equation} \label{omega1}
1=\Omega_{m}\left( 1 - \frac{l_{p}^{2}\gamma_{q}}{R^{2}}\right)+\Omega_{\Lambda }
\end{equation}
We could also define a new $\rho_{crit}$, namely
\begin{equation}
\tilde{\rho}_{crit}=\frac{\rho_{crit}}{1-\frac{l_{p}^{2}\gamma_{q}}{R^{2}}}\simeq \rho_{crit}\left(1+\frac{l_{p}^{2}\gamma_{q}}{R^{2}} \right)
\end{equation}
as well as  $\tilde{\Omega}_{m}=\frac{\rho}{\tilde{\rho}_{crit}}$. Then we simply have 
\begin{equation}
1=\tilde{\Omega}_{m}+\Omega_{\Lambda}
\end{equation}

If we assume the equation of state of radiation the conservation law gives us
\begin{equation}
\left(\frac{\rho}{\rho_{0}}\right)^{1/2}=\frac{1}{a^{2}}
\end{equation}
with $a=R/R_0$. Then it is easy to see that the first Friedmann equation becomes
\begin{equation}
H^{2}=\frac{8\pi G}{3}\rho \left(1-\beta' \frac{R_{0}^{2}}{R^{2}} \right)=\frac{8\pi G}{3}\rho \left(1-\beta' \frac{1}{a^{2}} \right)
=\frac{8\pi G}{2}\rho \left(1-\beta'\left(\frac{\rho}{\rho_{0}} \right)^{1/2} \right)
\end{equation}
with $\beta'=\gamma_q l_p^2/R_0^2$
In the case of positive $\gamma_q$ (positive $\beta'$) it would make sense to intrduce a critical density
\begin{equation}
\tilde{\rho}_{cr}=\frac{\rho_{0}}{(\beta')^{2}}=\rho_{0}\frac{R_{0}^{4}}{\gamma_{q}^{2}l_{p}^{4}}=\mbox{constant}
\end{equation}
such that $H=0$ when $\rho=\tilde{\rho}_{cr}$. Although we will make a detailed comparison with other models at the end of the paper
we notice already here that in loop quantum gravity the expression is similar, i.e.,  
\begin{equation}
H^{2}=\frac{8\pi G}{3}\rho \left(1-\frac{\rho}{\rho_{C}^{(LQG)}} \right)
\end{equation}
This does not imply that there is quantum bounce only if $\gamma_q$ is positive. Indeed, in the next section by solving explicitly
the Friedmann equations that even in the case of $\gamma_q < 0$ the universe has no singularity at $R=0$. Different scenarios are possible,
mostly depending on the sign of $\gamma_q$ and the equation of state. 

Some mathematical features of classical and quantum
universes are common. In the following steps we will briefly discuss two solutions of the standard Friedmann equations without quantum corrections.
First let us consider
a toy universe with $\Lambda=0$, $\gamma_q=0$ and $k=0$  filled with radiation. It is an easy excersise to show
that the solution to the Friedman equations for $a=R/R_0$ reads $(a^2-1)/2=\pm \tau\equiv \sqrt{8\pi G\rho_0/3}(t-t_0)$. 
The two branches correspond to 
\begin{eqnarray}
a_+&=&\sqrt{2\tau + 1}, \,\,   \tau > -1/2 \nonumber \\ 
a_-&=&\sqrt{1-2\tau}, \,\,  1/2  > \tau
\end{eqnarray}
with $a_+(0)=a_-(0)$. The branch $a_-$ is decreasing whereas
$a_+$ is growing in time (see figure 1). It would be incorrect to try to avoid the singualrity by gluing
the two branches at $\tau=0$ discarding the rest. This would lead to an ambiguity in the solution as we would have
four possible solutions. This tells us that we can only glue the two branches if we arrive
at a unique smooth solution.  Secondly, we take the radiation case with $\Lambda=0$, $\gamma_q=0$ and $k=1$. 
Due to 
$H^{2}=\frac{8\pi G}{3}\rho - \frac{k}{R^{2}}$ the Hubble constant can be zero, but this corresponds to a local maximum as we will see. Indeed,
the solutions are 
\begin{eqnarray}
a_{-}(\tau)&=&\sqrt{-\frac{3}{8\pi G}\frac{1}{\rho_{0}}\frac{\tau^{2}}{R_{0}^{2}}- \frac{2\tau}{R_{0}^{2}}\sqrt{R_{0}^{4}-\frac{3}{8\pi G}\frac{1}{\rho_{0}}R_{0}^{2}}+1}
\nonumber \\
a_{+}(\tau)&=&\sqrt{-\frac{3}{8\pi G}\frac{1}{\rho_{0}}\frac{\tau^{2}}{R_{0}^{2}}+ \frac{2\tau}{R_{0}^{2}}\sqrt{R_{0}^{4}-\frac{3}{8\pi G}\frac{1}{\rho_{0}}R_{0}^{2}}+1}
\end{eqnarray}
which we plotted in figure 1.
There is a restriction on $R_0$ in form
$R_{0}\geq \sqrt{\frac{3}{8\pi G}\frac{1}{\rho_{0}}}$ and on $R$ given us the position of the maximum of $a$. The latter
is $R \le R_{max}= \sqrt{\frac{8\pi G}{3}\rho_{0}}R_{0}^{2}$ making the Hubble parameter vanish.  In the case of quantum universes as derived in this paper
we will see that  $H=0$ will either indicate a local minimum or an absolute minimum.

\begin{figure}[ht!]
\centering
\includegraphics[width=140mm]{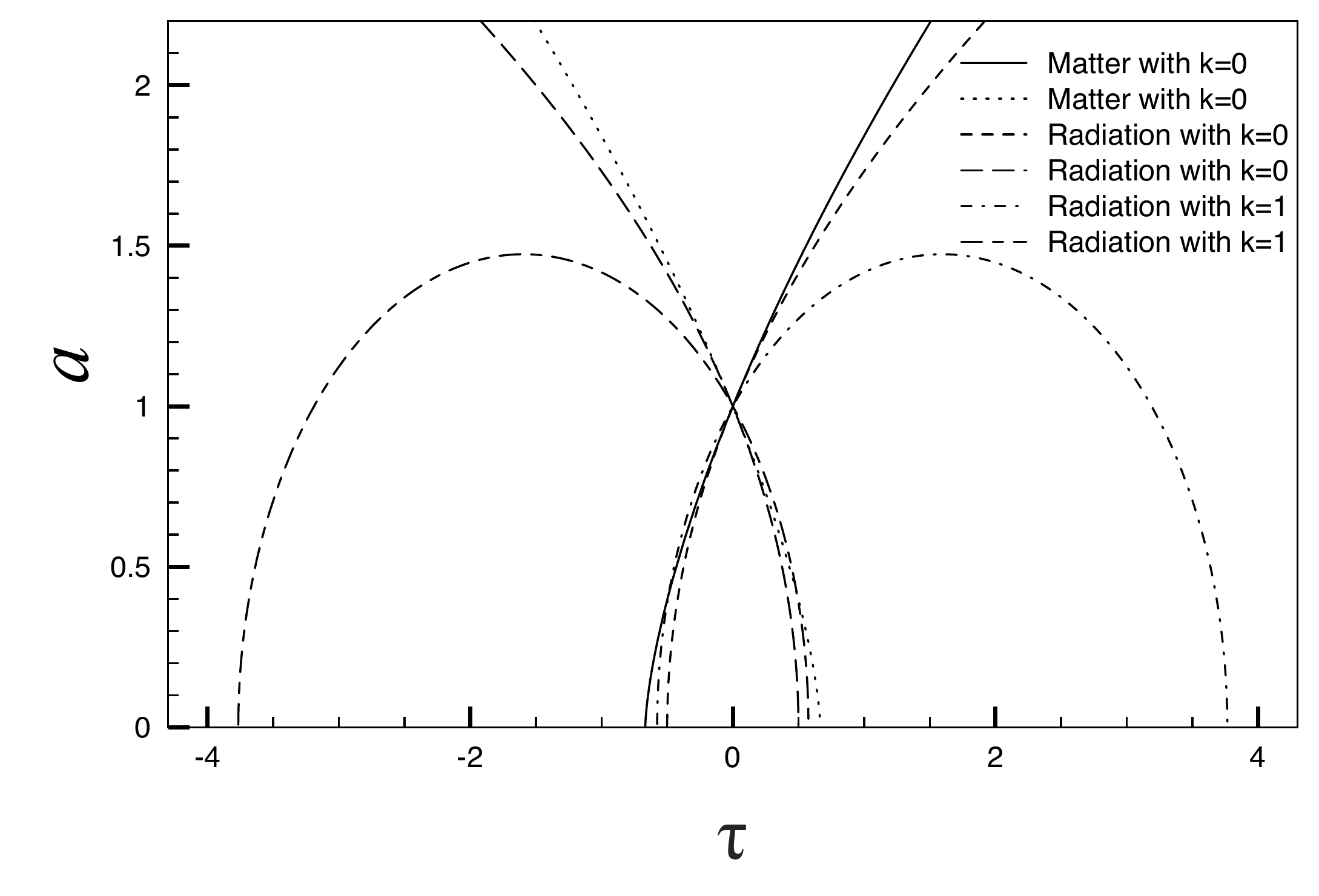}
\caption{Cosmological solutions for the scale parameter $a$ from the standard Friedmann equations.
See the text for more explanation.}
\end{figure}

\section{Newtonian Quantum Universes}
To find out the effect of the new term porportinal $\hbar$ in the Friedmann equations we start from the energy conservation equation
and use first an equation of state (EOS) of the form
\begin{equation}
p=(\gamma-1)\rho
\end{equation}
where $\gamma$ is not to be confused with $\gamma_{q}$. We can solve for $\rho$ in terms of $R$, namely
The standard solution is
\begin{equation}
\rho (R)=\rho_{0}\left(\frac{R_{0}}{R}\right)^{3\gamma}=\rho_0a^{-3\gamma}
\end{equation}
such that $\rho(R_{0})=\rho_{0}$. Inserting this into the first Friedmann equation with $k=0$, i.e., 
\begin{equation}
\left( \frac{dR}{dt}\right)^{2}=\frac{8\pi G}{3}\rho(R) R^{2}-\frac{8\pi G}{3}\rho(R) \beta
\end{equation}
with  $\beta= l_{p}^{2}\gamma_{q}$, we obtain
\begin{equation}
\left(\frac{dR}{dt} \right)^{2}=\frac{8\pi G}{3}\rho_{0}\frac{R_{0}^{3\gamma}}{R^{3\gamma}}\left[ R^{2}-\beta \right]
\end{equation}
In the integral form this reads as
\begin{equation}
t-t_{0}=\pm \frac{1}{R_{0}^{3\gamma/2}\sqrt{\frac{8\pi G}{3}\rho_{0}}} \int_{R_{0}}^{R}\frac{\bar{R}^{3\gamma/2}}{\sqrt{\bar{R}^{2}-\beta}}d\bar{R}
\end{equation}
The behavior of the solution depends strongly on the sign of $\beta'$ (which is the same as the
sign of $\gamma_q$) and on the equation of state ($\gamma$). It therefore make sense to discuss the different cases
separately.

\subsection{Case $\beta<0$}

\subsubsection{Radiation ($\gamma=4/3$)}
In this case the solution can be given in terms of standard fuctions, namely
\begin{equation}
\tau \equiv \sqrt{\frac{8\pi G}{3}}(t-t_{0})=\pm \frac{1}{2}\frac{R}{R_{0}^{2}}\sqrt{R^{2}+|\beta|}\mp \frac{1}{2}\frac{|\beta|}{R_{0}^{2}}\ln \left[R+\sqrt{R^{2}+|\beta|} \right]+D
\end{equation}
where D takes care of the initial value $R(t_0)=R_0$.
After implementing the initial value  we obtain
\begin{equation} \label{solution1}
\tau=\sqrt{\frac{8\pi G}{3}\rho_{0}}(t-t_{0})=\pm \frac{1}{2}\left[ \sqrt{a^{2}+\beta'}-\sqrt{1+\beta'}\right] \mp\frac{1}{2}\beta' \ln \left[ \frac{a+\sqrt{a^{2}+\beta'}}{1+\sqrt{1+\beta'}}\right]
\end{equation}

\begin{figure}[ht!]
\centering
\includegraphics[width=140mm]{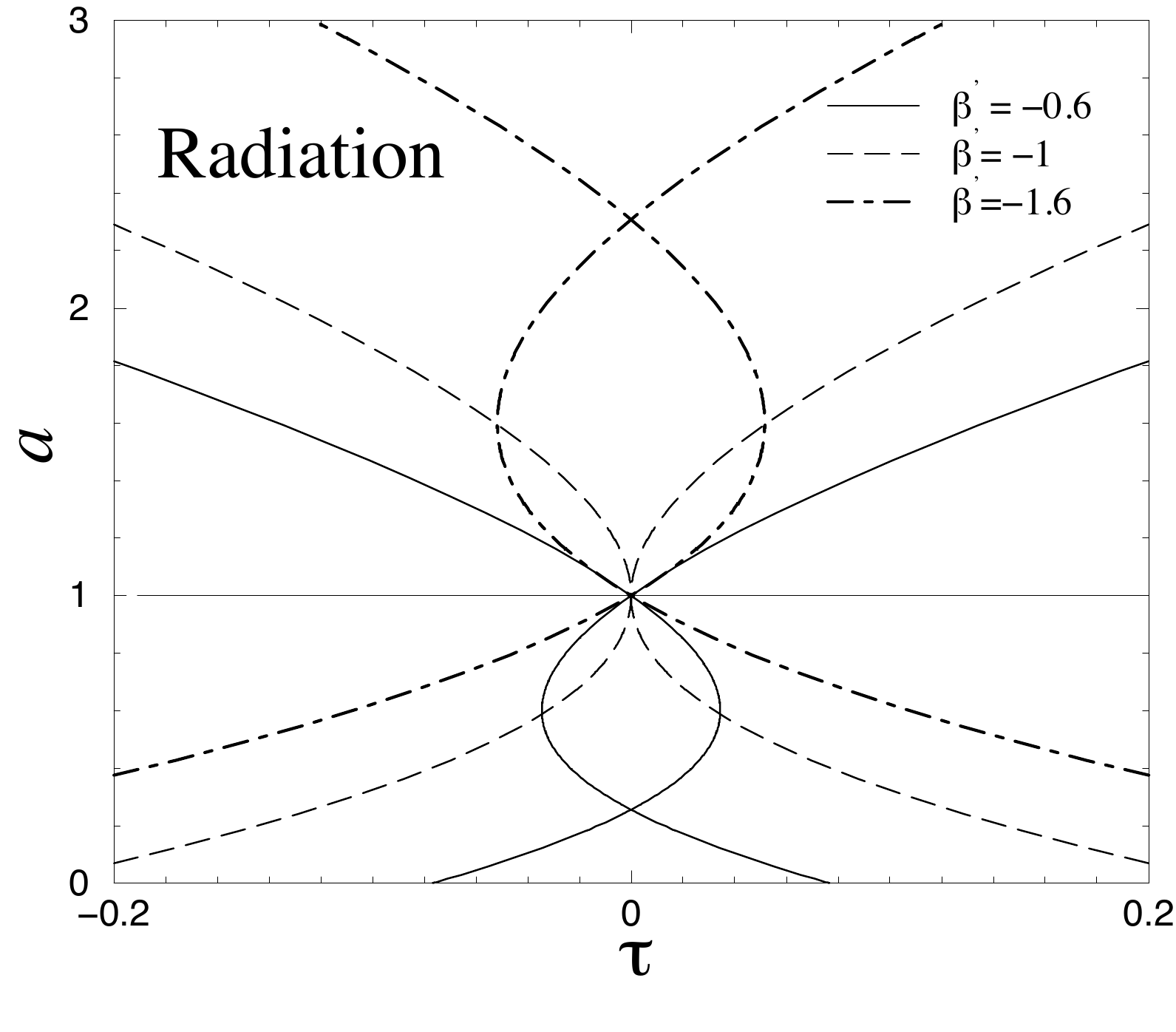}
\caption{Cosmological solutions for the scale parameter $a$ from the modified Friedmann equations.
This figure displays the solution for the radiation equation of state and negative $\beta'$. See the text for more explanation.}
\end{figure}
\begin{figure}[ht!]
\centering
\includegraphics[width=140mm]{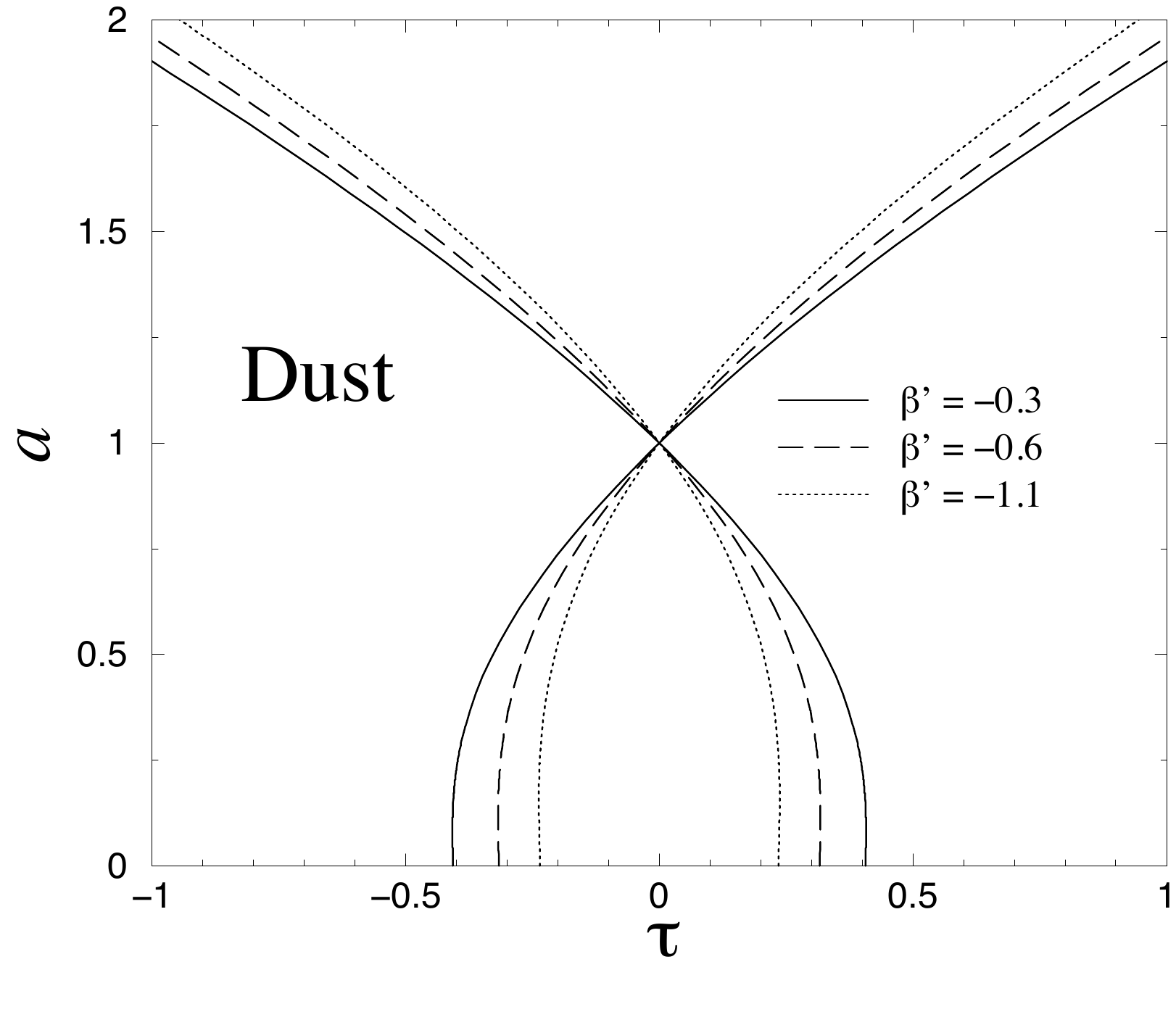}
\caption{The same as in figure 2, but for the case of dust. See text for a detailed discussion.}
\end{figure}

Figure 2 shows the solutions according to equation (\ref{solution1}). To the right and left of the straight line
$\tau=0$ we have the different branches due to the $\pm$ signs in (\ref{solution1}). As long as $|\beta'| < 1$ we 
get always a non-singular universe: the expanding universe  starts at a non-zero value $R_{min} < R_0$ (below the line $a=1$) 
determined by the single-valuedness of $a$. The mirror universe in such a case is a contracting one ending at the same $R_{min}$
This is shown in the figure 2 for $\beta'=-0.6$. 
For $|\beta'| > 1$ we obtain a singular universe starting at $R=0$ and ending at some $R_{max} > R_0$ following one of the signs
(the other sign gave the mirror collapsing universe starting at $R_{max}$ and contracting to zero.  The critical point seems to be
$\beta'=-1$. If we choose the solution according to one sign and the single-valuedness of the solution
we would end up with two expanding universes, one starting from zero and expanding up to $R_0$, the other starting from
$R_0$ and expanding up to infinity (a similar picture emerges for the collapsing branch). However, these two
curves merge smoothly at $R_0$ and therefore we can construct a unique forever expanding universe starting at zero (and similarly the
mirror image). We conclude that for negative $\beta'$ with $|\beta'| < 1$ the quantum effect is that the universe starts at a finite
value of $R_0$. In section V we solve these equations by including the cosmological constant. The next section
is devoted to the comparison with other quantum models. In the the last section we  draw our conclusions.

\subsubsection{Dust ($\gamma=1$)}
The integral to solve is now
\begin{equation}
t-t_{0}=\pm \frac{1}{\sqrt{\frac{8\pi G}{3}\rho_{0}R_{0}^{3}}}\int_{R_{0}}^{R}dR\sqrt{\frac{\bar{R}^{3}}{\bar{R}^{2}+|\beta|}}
\end{equation}
which can be rewritten in terms of $a=R/R_{0}$ as:
\begin{equation}\label{ii}
\sqrt{\frac{8\pi G}{3}\rho_{0}}(t-t_{0})=\pm\mathcal{I}(a),\quad\mathcal{I}(a)=\int_{1}^{a}d\tau \sqrt{\frac{\tau^{3}}{\tau^{2}+|\beta^{'}|}},\quad\beta^{'}=\frac{\beta}{R_0^2}.
\end{equation}
In order to solve the integral appearing in the above expression we first rewrite it as follows:
\begin{equation}
\mathcal{I}(a)=\int_{1}^{a}d\tau \frac{\tau^{2}}{\sqrt{\tau(\tau^{2}+|\beta^{'}|)}}.
\end{equation}
Using 230.1 in \cite{Byrd} yields:
\begin{equation}
\mathcal{I}(a)=\frac{1}{3}\left[2\sqrt{a(a^2+|\beta^{'}|)}-2\sqrt{1+|\beta^{'}|}-|\beta^{'}|\int_1^a\frac{d\tau}{\sqrt{\tau(\tau^{2}+|\beta^{'}|)}}\right].
\end{equation}
At this point, it is convenient to split the integral above as
\begin{equation}
\int_1^a\frac{d\tau}{\sqrt{\tau(\tau^{2}+|\beta^{'}|)}}=\int_0^a\frac{d\tau}{\sqrt{\tau(\tau^{2}+|\beta^{'}|)}}-\int_0^1\frac{d\tau}{\sqrt{\tau(\tau^{2}+|\beta^{'}|)}}.
\end{equation}
Both integrals on the r.h.s. of the above expression can be computed by means of 239.00 in \cite{Byrd} and we find that
\begin{equation}
\int_1^a\frac{d\tau}{\sqrt{\tau(\tau^{2}+|\beta^{'}|)}}=\frac{1}{|\beta^{'}|^{1/4}}\left[F\left(\cos^{-1}\left(\frac{|\beta^{'}|^{1/2}-a}{|\beta^{'}|^{1/2}+a}\right),\frac{1}{\sqrt{2}}\right)-F\left(\cos^{-1}\left(\frac{|\beta^{'}|^{1/2}-1}{|\beta^{'}|^{1/2}+1}\right),\frac{1}{\sqrt{2}}\right)
\right],
\end{equation}
where $F(\varphi,k)$ denotes the elliptic integral of the first kind with amplitude and modulus represented by $\varphi$ and $k$, respectively. Finally, we obtain
\begin{equation}\label{tt}
\mathcal{I}(a)=\frac{2}{3}\left[\sqrt{a(a^2+|\beta^{'}|)}-\sqrt{1+|\beta^{'}|}\right]-\frac{|\beta^{'}|^{3/4}}{3}\left[F\left(\cos^{-1}\left(\frac{|\beta^{'}|^{1/2}-a}{|\beta^{'}|^{1/2}+a}\right),\frac{1}{\sqrt{2}}\right)-F\left(\cos^{-1}\left(\frac{|\beta^{'}|^{1/2}-1}{|\beta^{'}|^{1/2}+1}\right),\frac{1}{\sqrt{2}}\right)
\right].
\end{equation}
The results are plotted in figure 3. 
Following one branch, i.e. one sign, we conclude that all universe are singular as they start at zero (or end at zero). We conclude that in order to get
a non-singular universe in the case of negative $\beta'$ the equation of state plays a crucial role.   
Needless to say that at the beginning of the universe a relativistic equation of state is preferred.

\subsection{Case $\beta>0$}
\subsubsection{Radiation ($\gamma=4/3$)}
The solution is now given by
\begin{equation}
\sqrt{\frac{8\pi G}{3}\rho_{0}}(t-t_{0})=\pm \frac{1}{2}\frac{R}{R_{0}^{2}}\sqrt{R^{2}-\beta}\pm\frac{1}{2}\frac{\beta}{R_{0}^{2}} \ln \left[R+\sqrt{R^{2}-\beta} \right]+C
\end{equation}
where $C$ is a constant. In terms of $a=R/R_{0}$ and implementing the initial value explicitly it reads
\begin{equation} \label{sol}
\tau=\sqrt{\frac{8\pi G}{3}\rho_{0}}(t-t_{0})=\pm\frac{1}{2}\left[ \sqrt{a^{2}-\beta'}-\sqrt{1-\beta'}\right] \pm \frac{1}{2}\beta' \ln \left[\frac{a+\sqrt{a-\beta'}}{1+\sqrt{1-\beta'}} \right]
\end{equation}
We see that in general the case with $\gamma_{q}>0$ imposes a certain limit upon the value of $R$,
namely $R^{2}\geq R^2_{min}\equiv \beta$ or, equivalently, $a^{2}\geq\beta'$.
This is clearly reflected in figure 4 where we have plotted the solutions. 
\begin{figure}[ht!]
\centering
\includegraphics[width=140mm]{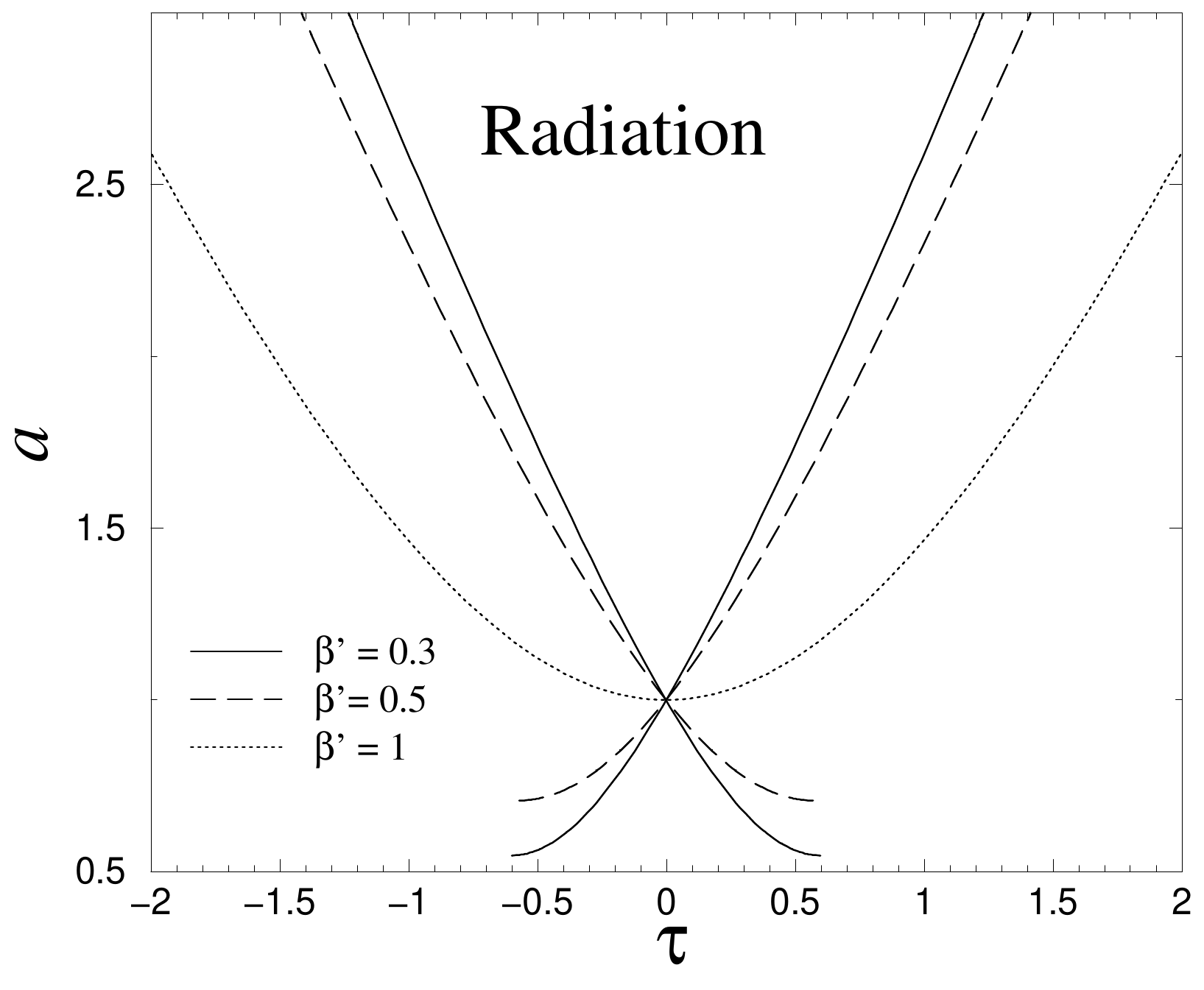}
\caption{Cosmological solutions for the scale parameter $a$ from the modified Friedmann equations.
This figure displays the solution for the radiation equation of state and positive $\beta'$. See the text for more explanation.}
\end{figure}
Since according to (\ref{sol}) we obtain a solution if $\beta' \le 1$  we note that all universes start (or end) at $R_{min}$
as expected as long as $\beta'$ is smaller than one. In the case that $\beta'=1$ $R_{min}$ is the position of the local minimum.
This minimum joins the two branches with different signs smoothly and gives a unique solution. This universe is then different from the others
as it ``comes'' from infinity, reaches a minimum and expands again.

\subsubsection{Dust ($\gamma=1$)}
We end up with the computation of the following integral
\begin{equation}\label{ee1}
\tau=\sqrt{\frac{8}{3}\pi G\rho_0}(t-t_0)=\pm I(a),\quad I(a)=\int_1^a d\tau\frac{\tau^2}{\sqrt{\tau(\tau^2-\beta^{'})}}.
\end{equation}
If we apply 230.1 form \cite{Byrd} to $I(a)$, we find that
\begin{equation}
I(a)=\frac{1}{3}\left[2\sqrt{a(a^2-\beta^{'})}-2\sqrt{1-\beta^{'}}+\beta^{'}\int_1^a \frac{d\tau}{\sqrt{\tau(\tau^2-\beta^{'})}}\right].
\end{equation}
Since the integrand is real, we must require that $0<\beta^{'} \le 1$. It is convenient to rewrite the integral above as follows
\begin{equation}
\int_1^a \frac{d\tau}{\sqrt{\tau(\tau^2-\beta^{'})}}=\int_{\sqrt{\beta^{'}}}^a \frac{d\tau}{\sqrt{\tau(\tau^2-\beta^{'})}}-\int_{\sqrt{\beta^{'}}}^1 \frac{d\tau}{\sqrt{\tau(\tau^2-\beta^{'})}}.
\end{equation}
The integrals appearing on the r.h.s. of the above expression can be evaluated by means of 237.00 in \cite{Byrd} and we obtain
\begin{equation}
\int_1^a \frac{d\tau}{\sqrt{\tau(\tau^2-\beta^{'})}}=\frac{\sqrt{2}}{(\beta^{'})^{1/4}}\left[
F\left(\sin^{-1}{\sqrt{\frac{a-\sqrt{\beta^{'}}}{a}}},\frac{1}{\sqrt{2}}\right)-F\left(\sin^{-1}{\sqrt{1-\sqrt{\beta^{'}}}},\frac{1}{\sqrt{2}}\right)
\right].
\end{equation}
where $F$ is the elliptic integral of the first kind.
Hence, the integral $I(a)$ can be computed to be
\begin{equation}\label{ii1}
I(a)=\frac{2}{3}\left[\sqrt{a(a^2-\beta^{'})}-\sqrt{1-\beta^{'}}\right]+\frac{\sqrt{2}}{3}(\beta^{'})^{3/4}\left[
F\left(\sin^{-1}{\sqrt{\frac{a-\sqrt{\beta^{'}}}{a}}},\frac{1}{\sqrt{2}}\right)-F\left(\sin^{-1}{\sqrt{1-\sqrt{\beta^{'}}}},\frac{1}{\sqrt{2}}\right)
\right].
\end{equation}
The results are presented in figure 5. In the case of positive $\beta'$ there is not much difference if we change the equation of state.
Therefore, the interpretations are similar to the radiation case and $\beta'=1$ is again a special case. 
\begin{figure}[ht!]
\centering
\includegraphics[width=140mm]{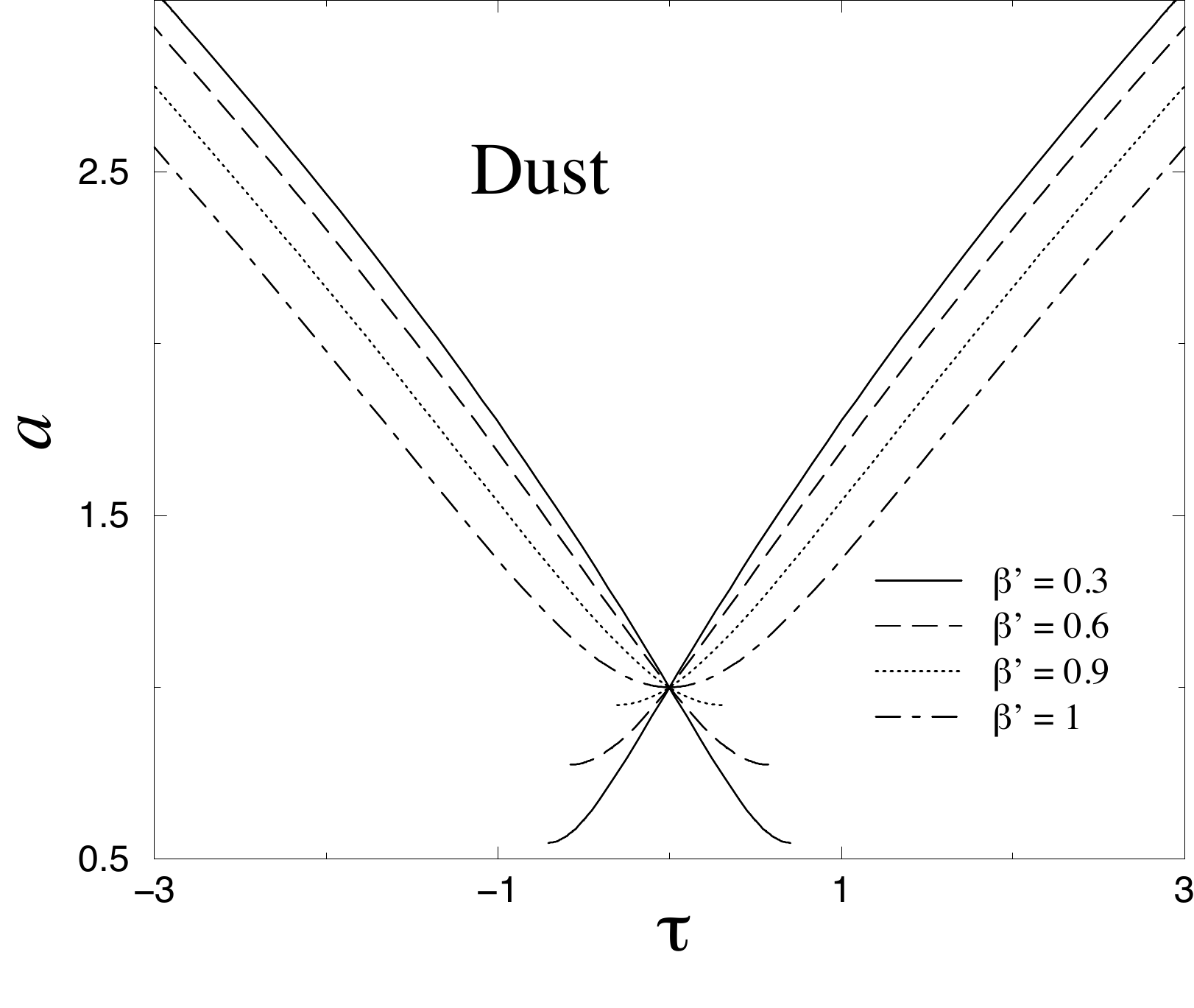}
\caption{Cosmological solutions for the scale parameter $a$ from the modified Friedmann equations as in figure 4, but for dust.
See the text for more explanation.}
\end{figure}

\section{The case with a Cosmological Constant} 
It is of some interest to treat the full Friedmann equation with the quantum corrections
and spatial flatness.
Including the cosmological constant $\Lambda$, the Friedmann equations read 
\begin{eqnarray} 
H^{2} &=&\frac{8}{3}\pi G \rho +\frac{1}{3}\Lambda-\frac{8}{3}\pi G \rho \frac{l_{p}^{2}\gamma_{q}}{R^{2}} \nonumber \\
\frac{d^{2}R}{dt^{2}}&=&-\frac{4\pi G}{3}\left( \rho+3p\right)R+\frac{1}{3}\Lambda R + 4\pi G l_{p}^{2}\gamma_{q} \frac{(\rho+p)}{R}
\end{eqnarray}
We will solve this case perturbatively.
\subsubsection{Radiation($\gamma=4/3$)} 
The integral to be solved in the radiation case and non-zero positive cosmological constant is
\begin{equation}
t-t_{0}=\pm \frac{1}{R_{0}^{2}}\int_{R_0}^R \frac{R^{2}dR}{\sqrt{ \frac{8\pi G}{3}\rho_{0}(R^{2}-\beta)+\frac{1}{3}\Lambda \frac{R^{6}}{R_{0}^{4}}}}
\end{equation}
which we can also rewrite this in terms of $a(t)$ and $\rho_{vac}\equiv\frac{\Lambda}{8\pi G}$ as follows
\begin{equation}\label{Lambda1}
\sqrt{\frac{8\pi G}{3}\rho_{0}}(t-t_{0})=\pm\mathcal{I}(a),\quad\mathcal{I}(a)=\int_1^a d\tau f(\tau),\quad f(\tau)=\frac{\tau^{2}}{\sqrt{\tau^{2}-\beta^{'}+\epsilon \tau^{6} }},\quad\epsilon=\frac{\rho_{vac}}{\rho_{0}}.
\end{equation}
Since $\epsilon\ll 1$, the integrand appearing in $I(a)$ can be expanded in powers of the small parameter $\epsilon$. Hence, we have
\begin{equation}
f(\tau)=\frac{\tau^2}{\sqrt{\tau^2-\beta^{'}}}-\frac{\tau^8}{2(\tau^2-\beta^{'})^{3/2}}\epsilon+\mathcal{O}(\epsilon^2).
\end{equation}
Taking into account that
\begin{equation}
F_0(a)=\int_1^a d\tau \frac{\tau^2}{\sqrt{\tau^2-\beta^{'}}}=\frac{\beta^{'}}{2}\ln{\frac{a+\sqrt{a^2-\beta^{'}}}{1+\sqrt{1-\beta^{'}}}}+\frac{a}{2}\sqrt{a^2-\beta^{'}}-\frac{1}{2}\sqrt{1-\beta^{'}}
\end{equation}
and
\[
F_1(a)=-\frac{1}{2}\int_1^a d\tau\frac{\tau^8}{(\tau^2-\beta^{'})^{3/2}}=\frac{35}{32}(\beta^{'})^3\ln{\frac{1+\sqrt{1-\beta^{'}}}{a+\sqrt{a^2-\beta^{'}}}}-\frac{a}{4\sqrt{a^2-\beta^{'}}}\left[\frac{a^6}{3}+\frac{7}{12}\beta^{'}a^4+\frac{35}{24}(\beta^{'})^2 a^2-\frac{35}{8}(\beta^{'})^3\right]
\]
\begin{equation}
-\frac{1}{4\sqrt{1-\beta^{'}}}\left[\frac{35}{8}(\beta^{'})^3-\frac{35}{24}(\beta^{'})^2-\frac{7}{12}\beta^{'}-\frac{1}{3}\right],
\end{equation}
we find that $a(t)$ is given at the first order in $\epsilon$ by the following expression
\begin{equation}
\tau=\sqrt{\frac{8\pi G}{3}\rho_{0}}(t-t_{0})=\pm\left[F_0(a)+F_1(a)\epsilon\right]+\mathcal{O}(\epsilon^2).
\end{equation}

\subsubsection{Dust ($\gamma=1$)}
In this case the integral to be solved has the form
\begin{equation}
\sqrt{\frac{8\pi G}{3}\rho_{0}}(t-t_0)=\pm S(a),\quad S(a)=\int_{1}^{a}d\tau g(\tau),\quad g(\tau)=\sqrt{\frac{\tau^{3}}{\tau^{2}-\beta^{'}+\epsilon \tau^{5}}}\quad\epsilon=\frac{\rho_{vac}}{\rho_{0}}.
\end{equation}
Expanding the integrand in powers of the small parameter $\epsilon$ yields
\begin{equation}
g(\tau)=\sqrt{\frac{\tau^3}{\tau^2-\beta^{'}}}-\frac{\tau^5}{2(\tau^2-\beta^{'})}\sqrt{\frac{\tau^3}{\tau^2-\beta^{'}}}\epsilon+\mathcal{O}(\epsilon^2).
\end{equation}
First of all, observe that
\begin{equation}
\int_1^a d\tau\sqrt{\frac{\tau^3}{\tau^2-\beta^{'}}}=\int_1^a d\tau\frac{\tau^2}{\sqrt{\tau(\tau^2-\beta^{'})}}=I(a)
\end{equation}
with $I(a)$ given by (\ref{ii1}). Let
\[
G(a)=-\frac{1}{2}\int_1^a d\tau\frac{\tau^5}{(\tau^2-\beta^{'})}\sqrt{\frac{\tau^3}{\tau^2-\beta^{'}}}=
-\frac{1}{2}\int_1^a d\tau\frac{\tau^7}{(\tau^2-\beta^{'})\sqrt{\tau(\tau^2-\beta^{'})}}
\]
\begin{equation}
=-\frac{1}{2}\int_1^a \frac{d\tau}{\sqrt{\tau(\tau^2-\beta^{'})}}\left[\tau^5+\beta^{'}\tau^3+(\beta^{'})^2\tau+\frac{(\beta^{'})^3}{2(\tau-\sqrt{\beta^{'}})}+\frac{(\beta^{'})^3}{2(\tau+\sqrt{\beta^{'}})}\right].
\end{equation}
Then the solution can be written as
\begin{equation}
\sqrt{\frac{8\pi G}{3}\rho_0}(t-t_0)=\pm\left[I(a)+G(a)\epsilon\right]+\mathcal{O}(\epsilon^2)
\end{equation}
The way to compute $G$ is long and we give all the details in the appendix.

\section{Comparison With Other Models}
\subsection{Loop quantum gravity and Friedmann equations}
For this section we will follow closely Refs. \cite{Gambinibook,Ashtekar2011}.
Let us consider the mini-superspace approach to Classical General Relativity for the $k=0$ case. After defining appropriate Ashtekar variables,
$c$ and $p$ \footnote{$a^2=p$ and $c=\dot a$}, which inherit the Poisson bracket given by $\{c,p\}=\frac{8}{3}\pi G \beta_{BI}$ ($\beta_{BI}$ is the Barbero--Immirzi parameter \footnote{The value
$\beta_{BI} \approx 0.2375$, as suggested by black hole physics, will be considered along the rest of the manuscript} and $G$ is Newton's
constant), the gravitational Hamiltonian constraint acquires the usual form
\begin{equation}
\mathcal{H}_{G} = -\frac{6}{\beta_{BI}^{2}}c^2\sqrt{|p|}.
\end{equation}
The contribution for a massless and free scalar field with a Hamiltonian constraint given by
\begin{equation}
\mathcal{H}_{\phi} = 8 \pi G \frac{p_{\phi}^2}{|p|^{3/2}}.
\end{equation}
Therefore, defining the Hubble parameter as $H=\dot p /(2 p)$ and the matter density for the scalar field
as $\rho = p_{\phi}^2/(2|p|^3)$, we get the total Hamiltonian constraint as $16 \pi G\left(\mathcal{H}_{G}+\mathcal{H}_{\phi}\right)$,
from which the usual Friedmann equation,
\begin{equation}
H^2= \frac{8 \pi G}{3} \rho,
\end{equation}
which predicts the usual big--bang singularity (the volume of the Universe goes to zero at $t=0$), can be recovered.

To proceed with quantization we have to promote $\mathcal{H}_{G}$ to a quantum operator. The imposibility lies in the fact that there is not quantum operator
associated to $c$. The usual way to circunvent this problem is called polymerization (see the \cite{Ashtekar2011} and references therein for technical details
on the procedure).

The important point is that the equations of motion derived from certain $\mathcal{H}_{\textrm{eff}}$,
given by $\dot p=\{p,\mathcal{H}_{\textrm{eff}}\}$, can be
expressed as a modified Friedmann equation, in the form
\begin{equation}
\label{lqgfriedmann}
H^{2}=\frac{8\pi G}{3}\rho \left(1-\frac{\rho}{\rho_{C}^{(LQG)}} \right),
\end{equation}
where the critical density is given by
\begin{equation}
\rho_{C}^{(LQG)} =\frac{3}{8 \pi G \beta_{BI}^{2}\mu_{0}^2}\approx 0.41\, \rho_{p}.
\end{equation}
The key point is that, in essence, the modified Friedmann equation leads to a non-singular evolution. Moreover, $\dot a$ vanishes at $\rho_{\textrm{crit}}$
and the Universe bounces. In the limit $\mu_{0}\rightarrow 0$, which corresponds to $G \hbar \rightarrow 0$, the critical density becomes infinity and the
classical singularity appears.

At this point some comments are in order. First of all, let us recall that the $\hbar-$corrections to the Friedmann equation ifor the $k=0$ case can be written as
\begin{equation}
\label{our}
H^{2}=\frac{8\pi G}{3}\rho \left(1-\beta'\frac{1}{a^2}\right).
\end{equation}
Therefore, some similarities and differences with the LQG--corrected Friedmann equation, given by Eq. (\ref{lqgfriedmann}), are present.
For positive $\gamma_q$, as commented before, a constant critical density can be obtained within our model considering radiation, although in this case the Friedmann equation is not
completely similar to the LQG case. But qualitative similarities persist.  
For negative $\gamma_q$ given the different sign which appears in Eq. (\ref{our}) compared to that of Eq. (\ref{lqgfriedmann}), an immediate comparison between
both predictions for the critical density is not evident. Nevertheless we find a quantum effect for negative $\gamma_q$ as the expanding universe starts
at a finite nonzero value for $R$.  The curcial sign of $\gamma_q$ will get also reflected in comparison with models other than Loop Qunatum Gravity.

\subsection{The generalized uncertainty principle, Snyder--deformed algebra and Friedmann equations}
As we have briefly commented, quantum corrections to the Friedmann equation can be implemented by considering Planck--scale modifications to the Hamiltonian
constraint, which lies at the heart of LQG. However, a different approach can be considered. What is the effect, if there is any, of deforming the usual Poisson brackets structure instead of deforming the
Hamiltonian constraint.

Without introducing Ashtekar variables,
interestingly, the generalized uncertainty principle (GUP) provides
a theoretical framework where this deformation appears and a consequence of the existence of a minimum length \cite{Sabine2013}.

The starting point is the formulation of ordinary canonical dynamics in FRW geometries. This dynamics is summarized in
the scalar constraint
\begin{equation}
\label{constr}
\mathcal{H}=-\frac{2 \pi G}{3}\frac{p_{a}^{2}}{a}-\frac{3}{8 \pi G}a k +a^3 \rho=0.
\end{equation} 
Isotropy makes $\{a,p_{a}\}=1$ the only non-vanishing Poisson bracket.

The equations of motion (in particular the Hubble equation) are easily deduced from 
the scalar constraint Eq. (\ref{constr}) and from
$\dot a= \{a,\mathcal{H}_{E}\}$ and 
$\dot p_{a}=\{p_{a},\mathcal{H}_{E}\}$, where the extended Hamiltonian is given by
\begin{equation}
\mathcal{H}_{E}=\frac{2 \pi G}{3}N\frac{p_{a}^{2}}{a}+\frac{3}{8 \pi G}Na k -N a^3 \rho=0 +\lambda \Pi.
\end{equation}
Here, $N=N(t)$ is the lapse function, $\lambda$ is a Lagrange multiplier and $\Pi$ is the momenta conjugate to $N$.

In the GUP framework, up to the first order in the deformation parameter, $\alpha$, the new Poisson bracket is
$\{a,p_{a}\}=1-2 \alpha p_{a}$. Using the new Hamilton's equations and again Eq. (\ref{constr}), the GUP--corrected
Hubble equation acquires the form \cite{Ali2014}
\begin{equation}
H^2=\left(\frac{8 \pi G}{3}\rho -\frac{k}{a^2}\right)
\left[1-2\alpha a^2 \sqrt{\frac{3}{2\pi G}}\left(\rho-\frac{3}{8 \pi G}\frac{k}{a^2}\right)^{1/2}\right].
\end{equation}

In particular, for the flat case ($k=0$), the modified Hubble equation reads
\begin{equation}
\label{GUPFriedmann}
H^2=\frac{8 \pi G}{3}\rho
\left(1-2\alpha a^2 \sqrt{\frac{3}{2\pi G}}\rho^{1/2}\right).
\end{equation}
At this point, it is important to recall that GUP gives place to a minimum length which, is this case, 
and taking $\alpha>0$, is associated
with the scale factor, $a(t)$. Therefore the critical density given by
\begin{equation}
\rho_{c}=\frac{2\pi G}{12 \alpha^{2}a^4},	
\end{equation}
remains finite (this situation is reminiscent of the appearance of a remnant mass in the GUP case for $\alpha>0$).

At this point it is interesting to consider some specific models for matter in Eq. (\ref{GUPFriedmann}).

\begin{itemize}
\item radiation ($\gamma=4/3$): In this case, Eq. (\ref{GUPFriedmann}) reads $H^2=\frac{8 \pi G}{3}\rho
\left(1-2\alpha \sqrt{\frac{3}{2\pi G}}\rho_{0}^{1/2}\right)$ and the critical density is given by $\rho^{GUP}_{crit}=\pi G/6 \alpha^2$.
\item dust ($\gamma=1$): In this case, Eq. (\ref{GUPFriedmann}) reads $H^2=\frac{8 \pi G}{3}\rho
\left(1-2\alpha \sqrt{\frac{3}{2\pi G}}\rho_{0}^{2/3}\rho^{-1/6}\right)$ and the critical density is given by 
$\rho^{GUP}_{crit}=\left(\frac{3}{2\pi G}\right)^3(2\alpha)^6\rho_{0}^4$.
\end{itemize}

A similar result can be obtained by invoking Snyder's non-commutative space, which gives place also to a deformed
Heisenberg algebra. In particular, the authors of Ref. \cite{Marco2009}, after replacing the usual Poissonian structure
between $a$ and $p_{a}$ by $\{a,p_{a}\}=\sqrt{1-\alpha p_{a}^2}$, obtained the following modified Friedmann equation:
\begin{equation}
\label{SnyFriedmann}
H^2=\left(\frac{8 \pi G}{3}\rho -\frac{k}{a^2}\right)
\left[1-\frac{3\alpha}{2\pi G}a^2 \left(a^2\rho-\frac{3}{8 \pi G}k \right)\right].
\end{equation}
Again considering the flat case the authors deduce
\begin{equation}
H^2=\frac{8 \pi G}{3}\rho
\left(1-a^4\alpha\frac{\rho}{\rho_{c}}\right),
\end{equation}
where $\rho_{c}=\frac{2\pi G}{3 \alpha}\rho_{p}$. In this last step it is also assumed, as a consequence of the deformed
algebra, the existence of a minimum length. Let us note again that $\alpha>0$ is necessary to smooth out the
singularity.

After considering Eq. (\ref{SnyFriedmann}) for radiation and dust matter, we obtain

\begin{itemize}
\item radiation ($\gamma=4/3$): $H^2= \frac{8\pi G}{3}\rho \left(1-\frac{3\alpha}{2\pi G}\rho_{0}\right)$ and $\rho^{Sny}_{crit}=2 \pi G/3 \alpha$.
\item dust ($\gamma=1$): $H^2= \frac{8\pi G}{3}\rho \left(1-\frac{3\alpha}{2\pi G}\rho^{4/3}_{0}\rho^{-1/3}\right)$ and 
$\rho^{Sny}_{crit}=\left(\frac{3\alpha}{2\pi G}\right)^3 \rho_{0}^4$.
\end{itemize}

Apart from comparing the critical density of our model, both in the dust and radiation cases, with those presnt in the previously mentioned approaches, 
it would be also interesting to show if our modified Friedmann equation (in the flat case) can be expressed in any of the forms predicted by LQG, Snyder or 
GUP, for certain polytropic fluid.
\\
\\
Specifically, it can be shown that our modified (spatially flat) Friedmann equation corresponds to: 
\begin{itemize}
\item Snyder's Friedmann when $\gamma=2$
\item GUP's Friedmann when $\gamma=8/3$
\item LQG's Friedmann when $\gamma=2/3$
\end{itemize}

The plausibility of the matter content is usually adressed with the help of the energy conditions. Introducing the variable $\omega=\gamma-1$, the 
energy conditions corresponding to the fluids with an equation of state of the form $p=\omega \rho$, for which $\omega=c_{s}^2$ (the sound speed associated
with this equation of state), are \cite{Hydrbook}:
\begin{itemize}
\item weak: $\rho+p\ge 0$, $\rho\ge 0$ $\leftrightarrow$ $\omega \ge -1$
\item strong: $\rho+p\ge 0$, $\rho+3p\ge 0$ $\leftrightarrow$ $\omega \ge -1/3$
\item dominant: $\rho\ge p$ $\leftrightarrow$ $-1\le\omega \le 1$
\end{itemize}

In particular, our model reproduces Snyder's corrections to the Friedmann equation when $\omega=1$. This corresponds to a ultrastiff or incompressible 
fluid which has been proposed as a possible description of the very early universe \cite{Hydrbook}. Moreover, this fluid is equivalent to
a free masless scalar \cite{Cosmologybook}. For this fluid, all the energy conditions are satisfied. In case of dealing with a $\omega=5/3$ fluid, the GUP
case is reproduced. In this situation, the dominant condition is violated. Finally, our model reproduces the LQG--corrected Friedmann equation when
$\omega=-1/3$. Interestingly, again in this situation, which corresponds to certain dark--energy model \cite{Ellisbook}, 
all the energy conditions are satisfied.

\subsection{Entropy corrections and Friedmann equations}

In recent years, quantum corrections to the Bekenstein--Hawking entropy have been shown to be either logarithmic or power--law. 
While the first kind of corrections usually arises from a minimum length scenario (such as LQG, GUP, etc) (see \cite{Sabine2013} and references therein), 
the second one deals with the entanglement 
of quantum fields inside and outside the horizon \cite{Saurya2008}. Moreover, the deep connection between gravity and thermodynamics, reinforced by 
Jacobson \cite{Jacobson1995} and Padmanabhan \cite{Padma2010}, made some authors \cite{Radicella2010} derive modified Friedmann equations by using 
corrections to the entropy in addition with the ideas explored in Refs. \cite{Jacobson1995} and \cite{Padma2010}. In addition, the entropy approach
developed by Verlinde \cite{Verlinde2011} has been employed \cite{Hendi2011}, assuming power--law corrections to the entropy, to obtain corrections to Friedmann
equations. 

Although the authors of \cite{Radicella2010} derive corrections to the Friedmann equations, they depend on the detailed gravity--thermodynamics 
connection. Even more, their main result (regarding our work), which is for the flat case, can be expressed as their Eq. (10), which reads
\begin{equation}
\label{entropy}
H^2\left[(1+g(\alpha,H) \right]=\frac{8\pi G}{3}\rho,
\end{equation}
where $g$ is a complicated function of $\alpha$, which is either the parameter that goes with the log--correction (for instance, $\alpha=-1/2$ in LQG) or 
the power of the entropy correction, and $H$. 
However, in spite of the formal similitude between Eqs. (\ref{our}) and (\ref{entropy}), the dependence of Eq. (\ref{entropy}) on $H$ makes the comparison between
both approaches very difficult to establish, unless some specific matter contents are considered.

In the case of power--law entropic corrections \cite{Hendi2011}, the key point is to notice that the Newtonian force gets corrected as
\begin{equation}
F=-\frac{G M m}{R^2}\left[1-\frac{\alpha}{2}\left(\frac{r_{c}}{R}\right)^{\alpha-2} \right],
\end{equation}
where $r_{c}$ is some crossover scale model--dependent and $\alpha$ is, as in the previous case, the power of the entropy correction. For the flat case, the
authors of Ref. \cite{Hendi2011} obtain
\begin{equation}
\label{entropic}
H^2=\frac{8 \pi G}{3}\rho \left[1-\beta_{PL} \left(\frac{r_{c}}{R} \right)^{\alpha-2} \right],
\end{equation}
where, assuming again an equation of the state of the form $p = \omega \rho$, $\beta_{PL}$ is given by 
$\beta_{PL}=\frac{\alpha}{2}\frac{3 \omega +1}{3 \omega +\alpha -1}$. Therefore, in spite of the similarities, Eqs. (\ref{our}) and (\ref{entropic}) are not
equivalent under any circumstances ($\alpha=0 \rightarrow \beta_{PL}=0$).

\section{Conclusions}
In this paper we have attempted a quantum cosmology based on the quantum corrections to the Newtonian potential
and repeating the derivation of Friedmann equation within the Newtonian formalism. The latter is known
to reproduce the correct Friedmann equations. This is one or the reasons why we believe that the quantum corrected
equations might hint towards what one would call the full fledge quantum cosmology. Indeed, with a certain choice
of the sign of the quantum correction we qualitatively agree with other models of a quantum universe.
In such a case a collapsing universe bounces off a minimum length proportional to the Planck length and begins to
expand again. Other quantum effects, for the opposite sign of this correction, manifest themselves
in a spontanesouly created universe at non-zero scale factor again close to the Planck length.
We believe that, at least qualitatively, this results go in the right and expected direction.

\section*{Appendix: Evaluation of integrals}
\appendix*
\setcounter{equation}{0}
We need to compute five integrals. Employing 230.01 in \cite{Byrd} yields
\begin{equation}
\int_1^a d\tau\frac{\tau^5}{\sqrt{\tau(\tau^2-\beta^{'})}}=\frac{1}{9}\left[2a^3\sqrt{a(a^2-\beta^{'})}-2\sqrt{1-\beta^{'}}+7\beta^{'}\int_1^a d\tau\frac{\tau^3}{\sqrt{\tau(\tau^2-\beta^{'})}}\right]
\end{equation}
Applying 230.01 in \cite{Byrd} to the last integral in the above expression leads to
\begin{equation}
\int_1^a d\tau\frac{\tau^3}{\sqrt{\tau(\tau^2-\beta^{'})}}=\frac{1}{5}\left[2a\sqrt{a(a^2-\beta^{'})}-2\sqrt{1-\beta^{'}}+3\beta^{'}\int_1^a d\tau\frac{\tau}{\sqrt{\tau(\tau^2-\beta^{'})}}\right].
\end{equation}
Concerning the last integral appearing on the r.h.s. in the expression above, we rewrite it as follows
\begin{equation}
\int_1^a d\tau\frac{\tau}{\sqrt{\tau(\tau^2-\beta^{'})}}=\int_1^a d\tau\sqrt{\frac{\tau}{\tau^2-\beta^{'}}}=\int_{\sqrt{\beta^{'}}}^a d\tau\sqrt{\frac{\tau}{\tau^2-\beta^{'}}}-\int_{\sqrt{\beta^{'}}}^1 d\tau\sqrt{\frac{\tau}{\tau^2-\beta^{'}}}
\end{equation}
and by applying 237.04 in \cite{Byrd} we find that
\begin{equation}
\int_1^a d\tau\frac{\tau}{\sqrt{\tau(\tau^2-\beta^{'})}}=\sqrt{2}(\beta^{'})^{1/4}\left[\int_0^{u_1}du~\mbox{nc}^2 u-\int_0^{\widetilde{u}_1}du~\mbox{nc}^2 u\right].
\end{equation}
Here, $\mbox{nc} u=1/\mbox{cn} u$ where $\mbox{cn}$ is one of the Jacobian elliptic functions and the associated amplitudes and moduli are given by
\begin{equation}\label{uk}
\varphi=\mbox{am} u_1=\sin^{-1}{\sqrt{\frac{a-\sqrt{\beta^{'}}}{a}}},\quad
\widetilde{\varphi}=\mbox{am} \widetilde{u}_1=\sin^{-1}{\sqrt{1-\sqrt{\beta^{'}}}},\quad
k^2=\frac{1}{2}=\widetilde{k}^2.
\end{equation}
Invoking 313.02 in \cite{Byrd} we find that
\begin{equation}
\int du~\mbox{nc}^2 u=\frac{1}{\widehat{k^{'}}^2}\left[\widehat{k^{'}}^2 u-E(\widehat{\varphi},\widehat{k})+\mbox{dn}u\mbox{tn}u\right]
\end{equation}
where $E$ denotes the elliptic integral of the second kind, $\widehat{k^{'}}=\sqrt{1-\widehat{k}^2}$ is the complementary modulus, $\mbox{dn} u$ and $\mbox{tn} u=\mbox{sn} u/\mbox{cn} u$ are the Jacobi elliptic functions. Taking into account that 111.00 and 122.01 in \cite{Byrd} imply that $E(0,k)=0=E(0,\widetilde{k})$, $\mbox{dn} 0=1$, $\mbox{tn} 0=0$ and moreover $k=\widetilde{k}=1/\sqrt{2}$, we obtain
\begin{equation}
\int_0^{u_1}du~\mbox{nc}^2 u-\int_0^{\widetilde{u}_1}du~\mbox{nc}^2 u=u_1-\widetilde{u}_1-2\left[E(\varphi,k)-E(\widetilde{\varphi},k)\right]+2\left(\mbox{dn}u_1\mbox{tn}u_1-\mbox{dn}\widetilde{u}_1\mbox{tn}\widetilde{u}_1\right).
\end{equation}
On the other hand, 121.01 in \cite{Byrd} implies that $u_1=F(\varphi,k)$ and $\widetilde{u}_1=F(\widetilde{\varphi},k)$ with $F$ denoting the elliptic integral of the first kind. Moreover, 120.01 allows also to find that
\[
\mbox{dn}u_1=\sqrt{1-k^2\sin^2{\varphi}}=\sqrt{\frac{a+\sqrt{\beta^{'}}}{2a}},\quad
\mbox{dn}\widetilde{u}_1=\sqrt{1-k^2\sin^2{\widetilde{\varphi}}}=\sqrt{\frac{1+\sqrt{\beta^{'}}}{2}}.
\]
Furthermore, by means of 121.00 in \cite{Byrd} we obtain
\[
\mbox{tn}u_1=\sqrt{\frac{a-\sqrt{\beta^{'}}}{\sqrt{\beta^{'}}}},\quad
\mbox{tn}\widetilde{u}_1=\sqrt{\frac{1-\sqrt{\beta^{'}}}{\sqrt{\beta^{'}}}}.
\]
Hence, we conclude that
\begin{equation}
\int_0^{u_1}du~\mbox{nc}^2 u-\int_0^{\widetilde{u}_1}du~\mbox{nc}^2 u=F(\varphi,k)-F(\widetilde{\varphi},k)-2\left[E(\varphi,k)-E(\widetilde{\varphi},k)\right]+\frac{\sqrt{2}}{(\beta^{'})^{1/4}}\left(\sqrt{\frac{a^2-\beta^{'}}{a}}-\sqrt{1-\beta^{'}}\right).
\end{equation}
This implies that 
\begin{equation}\label{i1}
\int_1^a d\tau\frac{\tau}{\sqrt{\tau(\tau^2-\beta^{'})}}=2\left(\frac{\sqrt{a(a^2-\beta^{'})}}{a}-\sqrt{1-\beta^{'}}\right)+\sqrt{2}(\beta^{'})^{1/4}\left\{F(\varphi,k)-F(\widetilde{\varphi},k)-2\left[E(\varphi,k)-E(\widetilde{\varphi},k)\right]\right\}
\end{equation}
and it is straightforward to verify that
\[
\int_1^a d\tau\frac{\tau^3}{\sqrt{\tau(\tau^2-\beta^{'})}}=\frac{2(a^2+3\beta^{'})}{5a}\sqrt{a(a^2-\beta^{'})}-\frac{2}{5}(1+3\beta^{'})\sqrt{1-\beta^{'}}
\]
\begin{equation}\label{i2}
+\frac{3}{5}\sqrt{2}(\beta^{'})^{5/4}\left\{F(\varphi,k)-F(\widetilde{\varphi},k)-2\left[E(\varphi,k)-E(\widetilde{\varphi},k)\right]\right\}.
\end{equation}
Finally, we find that
\[
\int_1^a d\tau\frac{\tau^5}{\sqrt{\tau(\tau^2-\beta^{'})}}=\frac{10 a^4+14\beta^{'}(a^2+3\beta^{'})}{45a}\sqrt{a(a^2-\beta^{'})}-\frac{10+14\beta^{'}(1+3\beta^{'})}{45}\sqrt{1-\beta^{'}}+
\]
\begin{equation}\label{i3}
\frac{7}{15}\sqrt{2}(\beta^{'})^{9/4}\left\{F(\varphi,k)-F(\widetilde{\varphi},k)-2\left[E(\varphi,k)-E(\widetilde{\varphi},k)\right]\right\}.
\end{equation}
Moreover, using 230.03 in \cite{Byrd} yields
\begin{equation}
\int_1^a\frac{d\tau}{(\tau-\sqrt{\beta^{'}})\sqrt{\tau(\tau^2-\beta^{'})}}=\frac{1}{2\beta^{'}}\left[2\sqrt{1-\beta^{'}}-2\sqrt{a(a^2-\beta^{'}
)}+\int_1^a d\tau\frac{\tau-\sqrt{\beta^{'}}}{\sqrt{\tau(\tau^2-\beta^{'})}}\right].
\end{equation}
Rewriting the integral appearing in the r.h.s. of the above expression as
\begin{equation}
\int_1^a d\tau\frac{\tau-\sqrt{\beta^{'}}}{\sqrt{\tau(\tau^2-\beta^{'})}}=\int_{\sqrt{\beta^{'}}}^a d\tau\sqrt{\frac{\tau-\sqrt{\beta^{'}}}{t(\tau+\sqrt{\beta^{'}})}}-\int^1_{\sqrt{\beta^{'}}} d\tau\sqrt{\frac{\tau-\sqrt{\beta^{'}}}{t(\tau+\sqrt{\beta^{'}})}}
\end{equation}
and applying 237.03 in \cite{Byrd} lead to
\begin{equation}
\int_1^a\frac{d\tau}{(\tau-\sqrt{\beta^{'}})\sqrt{\tau(\tau^2-\beta^{'})}}=\sqrt{2}(\beta^{'})^{1/4}\left[\int_0^{u_1}du~\mbox{tn}^2 u-\int_0^{\widetilde{u}_1}du~\mbox{tn}^2 u\right]
\end{equation}
with $u_1$ and $\widetilde{u}_1$ defined as in (\ref{uk}). By means of 316.02 in \cite{Byrd} we find that
\begin{equation}
\int_0^{u_1}du~\mbox{tn}^2 u-\int_0^{\widetilde{u}_1}du~\mbox{tn}^2 u=\frac{\sqrt{2}}{(\beta^{'})^{1/4}}\left(\frac{\sqrt{a(a^2-\beta^{'})}}{a}-\sqrt{1-\beta^{'}}\right)+2\left[E(\widetilde{\varphi},k)-E(\varphi,k)\right]
\end{equation}
and hence,
\begin{equation}
\int_1^a d\tau\frac{\tau-\sqrt{\beta^{'}}}{\sqrt{\tau(\tau^2-\beta^{'})}}=2\left(\frac{\sqrt{a(a^2-\beta^{'})}}{a}-\sqrt{1-\beta^{'}}\right)+2\sqrt{2}(\beta^{'})^{1/4}\left[E(\widetilde{\varphi},k)-E(\varphi,k)\right].
\end{equation}
Finally, we obtain that
\begin{equation}\label{i4}
\int_1^a\frac{d\tau}{(\tau-\sqrt{\beta^{'}})\sqrt{\tau(\tau^2-\beta^{'})}}=-(a-1)\frac{\sqrt{a(a^2-\beta^{'})}}{a\beta^{'}}+\frac{\sqrt{2}}{(\beta^{'})^{3/4}}\left[E(\widetilde{\varphi},k)-E(\varphi,k)\right].
\end{equation}
Observe that
\begin{equation}
\int_1^a\frac{d\tau}{(\tau+\sqrt{\beta^{'}})\sqrt{\tau(\tau^2-\beta^{'})}}=\int_{\sqrt{\beta^{'}}}^a\frac{d\tau}{(\tau+\sqrt{\beta^{'}})\sqrt{\tau(\tau^2-\beta^{'})}}-\int_{\sqrt{\beta^{'}}}^1\frac{d\tau}{(\tau+\sqrt{\beta^{'}})\sqrt{\tau(\tau^2-\beta^{'})}}.
\end{equation}
By means of 237.13 in \cite{Byrd} and taking into account that $\mbox{sn}0=0$ and $\mbox{cd} 0=1$ by 122.01 in \cite{Byrd}, we get
\begin{equation}
\int_1^a\frac{d\tau}{(\tau+\sqrt{\beta^{'}})\sqrt{\tau(\tau^2-\beta^{'})}}=\frac{\sqrt{2}}{(\beta^{'})^{3/4}}\left[u_1-\widetilde{u}_1+E(\widetilde{\varphi},k)-E(\varphi,k)+\frac{1}{2}\left(\mbox{sn}u_1\mbox{cd}u_1-\mbox{sn}\widetilde{u}_1\mbox{cd}\widetilde{u}_1\right)\right]
\end{equation} 
with $u_1$, $\widetilde{u}_1$ and $k$ given by (\ref{uk}). By means of the relation $\mbox{sn}^2 u+\mbox{cn}^2 u=1$, it is not difficult to verify that $\mbox{cn}u_1=(\beta^{'})^{1/4}/\sqrt{a}$ and $\mbox{cn}\widetilde{u}_1=(\beta^{'})^{1/4}$ and hence, we have
\begin{equation}
\mbox{cd} u_1=\frac{\mbox{cn}u_1}{\mbox{dn}u_1}=\frac{\sqrt{2}(\beta^{'})^{1/4}}{\sqrt{a+\sqrt{\beta^{'}}}},\quad
\mbox{cd} \widetilde{u}_1=\frac{\mbox{cn}\widetilde{u}_1}{\mbox{dn}\widetilde{u}_1}=\frac{\sqrt{2}(\beta^{'})^{1/4}}{\sqrt{1+\sqrt{\beta^{'}}}}.
\end{equation}
At this point it is straightforward to verify that
\begin{equation}\label{i5}
\int_1^a\frac{d\tau}{(\tau+\sqrt{\beta^{'}})\sqrt{\tau(\tau^2-\beta^{'})}}=\frac{1}{\sqrt{\beta^{'}}}\left(\frac{\sqrt{a(a^2-\beta^{'})}}{a(a+\sqrt{\beta^{'}})}-\frac{\sqrt{1-\beta^{'}}}{1+\sqrt{\beta^{'}}}\right)+\frac{\sqrt{2}}{(\beta^{'})^{3/4}}\left[F(\varphi,k)-F(\widetilde{\varphi},k)+E(\widetilde{\varphi},k)-E(\varphi,k)\right].
\end{equation}
With the help of (\ref{i1}), (\ref{i2}), (\ref{i3}), (\ref{i4}), and (\ref{i5}) we find that
\[
G(a)=-\frac{\sqrt{a(a^2-\beta^{'})}}{a}\left[\frac{20a^4+64\beta^{'}a^2-45(\beta^{'})^2a+417(\beta^{'})^2}{180}+\frac{(\beta^{'})^3}{4(a+\sqrt{\beta^{'}})}\right]+
\]
\[
\sqrt{1-\beta^{'}}\left[\frac{31}{15}(\beta^{'})^2+\frac{16}{45}\beta^{'}+\frac{1}{9}+\frac{(\beta^{'})^3}{4(a+\sqrt{\beta^{'}})}\right]-\frac{77}{60}\sqrt{2}(\beta^{'})^{9/4}\left\{F(\varphi,k)-F(\widetilde{\varphi},k)-2\left[E(\varphi,k)-E(\widetilde{\varphi},k)\right]\right\}.
\]
This completes the derivation.

\end{document}